# Application of gradient expansion to non-linear gravitational wave in plane-symmetric universe

Atsushi Taruya [*] and Yasusada Nambu [†]

Department of Physics, Nagoya University, Chikusa-ku, Nagoya 464-01, Japan

(October 6, 1995)

## Abstract

As an application of Gradient Expansion (long-wavelength) approximation, we studied the inhomogeneous universe including the gravitational wave(GW). For a plane-symmetric cosmological model, we could implement the 2nd order expansion and analyze the purely non-linear perturbation of GW. The result indicates that the non-linear effect plays an important role in anisotropic stage of the universe. We also confirm that the perturbation from the linear source term in gradient expansion is consistent with the result of linear perturbation.

## I. INTRODUCTION

Recently, there are greatly successful results in the measurement of the universe. The COBE's data [1] is one of the most important result because it indicates the temperature fluctuation of cosmic microwave background(CMB), which might be produced in the early stage of the universe. We can get the informations of the early universe by analyzing the fluctuation of CMB. We also have an alternative way of measurement of the early universe:

[*]e-mail address: ataruya@allegro.phys.nagoya-u.ac.jp

[†]e-mail address: nambu@allegro.phys.nagoya-u.ac.jp

using the gravitational wave (GW). The most advantageous point to use GW is that direct observation of early universe becomes possible because GW can propagate over the last scattering surface of the light. Concerning the detection of GW, the LIGO project has been developing [2]. Our knowledge of the early universe would become much wider from such a measurement in near future.

Taking into account of developments for these observations, theoretical study in the early universe based on general relativity becomes to play much important roles. A purpose of our study is to investigate the dynamical feature of inhomogeneous universe. Although it is well known from the observation that present our universe is quite homogeneous and isotropic, we cannot explain completely this fact. According to "Cosmic No-Hair Conjecture" [3], that is attractive idea, the present homogeneity was recovered from inhomogeneous spacetime by inflation. However, due to some counter-examples [4], we don't believe that this is true in general inhomogeneous case and need to seek the condition for the onset of inflation. The point that we should pay much attention is its non-linearity included in the Einstein equation.

There are three ways for analyzing the Einstein equation: searching exact solution, solving numerically and using approximation. When we want to investigate the feature of inhomogeneities in general case as possible, the approximation is an advantageous method unless its validity breaks down. "Gradient Expansion (GE)" method is one of such an analysis [5]- [12]. As is shown from its name, this is an approximation which treat the spatial derivative term as an expansion parameter. The most characteristic property is that one can include the non-linear derivative terms in the expansion parameters. This is the reason why one can investigate the non-linearity by using this method.

GE is valid when spatial derivative terms are smaller than other time derivative terms. Provided that $L$ is the scale of spatial inhomogeneity and $H$ is the scale of the time variation of the 3-metric $\gamma_{ij}$, this means

$$\frac{1}{a}|\partial_i \gamma_{jk}| \ll |\dot{\gamma}_{jk}| \iff H^{-1} \ll aL, \qquad (1.1)$$

where $a$ is scale-factor and $H \equiv \dot{a}/a$. Therefore GE method would become a good approximation if the scale of inhomogeneity is larger than the Hubble horizon size.

Concerning works on GE approximation, the 0th order solutions were investigated by Tomita [5]. These solutions were obtained by ignoring all spatial derivative in the Einstein equation and regarded as the most general ones in the sense that they include correct number of arbitrary spatial functions. Furthermore, he accomplished the formulation of the next order expansion keeping to include the right number of functions although it is not so applicable one. Note that the same formulation of the expansion scheme is recently investigated by Deruelle and Langlois [6]. According to the behavior of 0th order solutions, they become the form so-called quasi-isotropic expression, such as

$$\gamma_{ij} = a^2(t) h_{ij}(x^k). \tag{1.2}$$

In spite of ignoring some spatial functions, it is also 0th order solution in GE. Adopting this quasi-isotropic solution as the 0th order solution, Salopek and Stewart, Comer *et al.* studied about useful formulation for higher order expansion scheme [7] [8]: Salopek and Stewart built Hamilton-Jacobi theory for general relativity and solved the generating functional order by order. Comer *et al.* adopted the synchronous time slice and directly solved the Einstein equation expanded by spatial gradient. Both approaches are equivalent mathematically. Using these schemes, Croudace *et al.* [9] investigated the dust collapsing universe to clarify the relation with Zel'dovich approximation and found that one of the Szekeres solutions was reproduced as the 4th order solution of GE. Nambu and Taruya [10] analyzed the feature of inhomogeneous inflationary universe driven by minimally coupled inflaton field with arbitrary potential. Furthermore, Tomita and Deruelle studied non-linear behavior of perfect-fluid in cosmological case [11].

To consider the more generic situation, however, there are some remained issues. Constructing the cosmological model with the gravitational wave is one of the most important one. Although Salopek constructed the 0th order solution including the gravitational degree of freedom [12], no one investigates the 2nd order expansion. Due to the non-linearity of Einstein equation, the 2nd order correction of GE becomes to play an important role in the strong gravitational field. For example, colliding plane wave solutions which are the exact solutions of Einstein equation representing the collision of GW, shows us a lot of curious features of GW [13]. Hence, we expect that application of GE analysis for GW leads us to further recognition of the non-linear features.

However, one must keep in mind that the present expansion schemes can treat only some restricted inhomogeneities. It is due to adopting the quasi-isotropic condition as the 0th order solutions. In fact, the quasi-isotropic solutions are not general because some of the dynamics of gravitational field are neglected. We can show that a straightforward application of the expansion schemes cannot treat the GW completely.

In this paper, in order to clarify non-linear behavior of the GW in cosmological situations, we shall try to proceed to the 2nd order gradient expansion and analyze their evolution. Although in some restricted situations, we could complete this analysis without loss of feature of GW.

The paper is organized as follows. In Sec.II we make a short review about the expansion scheme of GE analysis and consider how to include GW. We construct a plane-symmetric cosmological model with GW and apply the GE analysis in Sec.III. The evolution equations up to the 2nd order gradient are derived and evaluated. In Sec.IV, we analyze our model using gauge-invariant linear perturbation. Comparing both analysis, we investigate the non-linear feature of GW. Summary and discussion are described in final section V.

## II. GRADIENT EXPANSION BY HJ METHOD

In order to investigate the cosmological model with GW, let us describe a short review of the expansion scheme based on Hamilton-Jacobi(HJ) method. Concerning the 2nd order expansion including GW, we shall clarify the difficulty in the scheme and consider how to avoid it.

HJ theory is one of the most applicable method for solving the various equations appeared in classical mechanics. For general relativity, HJ method is equivalent to solve Hamiltonian constraint and momentum constraint in terms of generating functional [7] [12]. GE analysis by HJ method is to expand the generating functional as follows:

$$\mathcal{S} = \mathcal{S}^{(0)} + \mathcal{S}^{(2)} + \cdots, \tag{2.1}$$

where upper script denotes the number of spatial derivative. In the case of the cosmological system coupled with n-scalar fields $\phi_A (A = 1, 2, \cdots n)$, we may take the following ansatz for the functional forms [7]:

$$\mathcal{S}^{(0)} = -\frac{2}{\kappa} \int d^3 x a^3 H(\phi_C), \tag{2.2}$$

$$\mathcal{S}^{(2)} = \int d^3 x a^3 \left[ J(\phi_C) R + K^{AB}(\phi_C) \phi_{A;m} \phi_B^{;m} \right], \quad (A, B, C = 1, 2, \cdots n), \tag{2.3}$$

where $a^3 \equiv \sqrt{det\{\gamma_{ij}\}}$. Because of their covariant forms with respect to spatial coordinate transformation, the momentum constraint is automatically satisfied order by order. We only have to pay attention to the Hamiltonian constraint. Solving Hamiltonian constraint order by order, we could get the solution of $H, K^{AB}$. In order to have the time evolution of the system, we substitute the generating functionals into the evolution equations. Under the synchronous time slice (take the gauge $N = 1, N_i = 0$), they become

$$\dot{\phi}_A = a^{-3} \frac{\delta \mathcal{S}}{\delta \phi_A}, \quad \dot{\gamma}_{ij} = \frac{4\kappa}{a^3} \frac{\delta \mathcal{S}}{\delta \gamma_{kl}} (\gamma_{ik} \gamma_{jl} - \frac{1}{2} \gamma_{ij} \gamma_{kl}). \tag{2.4}$$

By solving iteratively, one can obtain the solution in arbitrary order of accuracy.

However, when we consider the GW, the problem arises; the straightforward application of this scheme cannot treat GW. This impossibility comes from the ansatz of forms of the generating functionals. We can see that the 0th order functional (2.2) has only one dynamical degree of freedom of gravity, i.e, only conformal mode($a$) and no dynamical mode of GW. The 0th order evolution equations (2.4) are obtained by the variation of the functional (2.2),

$$\dot{\gamma}_{ij} = H \gamma_{ij}, \quad \dot{\phi}_A = -\frac{2}{\kappa} \frac{\partial H}{\partial \phi_A}. \tag{2.5}$$

Their solutions are expressed in the forms

$$\gamma_{ij}^{(0)} = a^2(\phi_A) h_{ij}(x), \quad \phi_A^{(0)} = \phi_A(t + t_0(x)),  \qquad (2.6)$$

where $h_{ij}(x)$ and $t_0(x)$ are arbitrary functions of the spatial coordinate. Since we ignore all spatial gradient in the 0th order expansion, the solutions (2.6) are equivalent to homogeneous flat FRW except for its spatial dependence. But due to lack of the GW mode in the 0th order solution, even though proceeding to the 2nd order, we cannot recover the dynamical feature of GW completely. Therefore we must require a bit device for the application of the scheme.

One of the easiest way to avoid such a problem is to consider a situation in which it is possible to include the GW by imposing some symmetry. In other words, we look for the system with GW which is equivalent to a cosmological model with conformal mode and the scalar fields. If it was possible, we could execute the expansion scheme and study about non-linear feature of GW.

## III. GRAVITATIONAL WAVE IN PLANE-SYMMETRIC COSMOLOGY

In previous section, we found that the present expansion scheme cannot be applicable directly to GW. In this section, we describe a cosmological model to which GE is applicable and implement the expansion up to the 2nd order. As is shown later, this model has two type of solutions concerning number of dynamical mode of GW. Here, we shall analyze only one solution(type A solution) for brevity. For the other solution(type B), we mention in Appendix C.

### A. Plane-symmetric cosmology

The main assumption to apply the scheme is to impose plane-symmetry, that is, there is spatial inhomogeneity only in the z-direction. Consider the cosmological model with the anisotropic metric

$$ds^2 = -dt^2 + a^2 \gamma_{ij} dx^i dx^j \quad ; \qquad (3.1)$$

$$\gamma_{ij} = \begin{pmatrix} (\xi^2 + \eta^2)/\eta & \xi/\eta & 0 \\ \xi/\eta & 1/\eta & 0 \\ 0 & 0 & 1 \end{pmatrix}. \tag{3.2}$$

Note that the dynamical variables $a, \xi, \eta$ are function of $(z, t)$ due to plane-symmetry. When we introduce a cosmological constant $\Lambda$ in this system, Lagrangian density takes the form as

$$\mathcal{L}_G = \frac{a^3}{2\kappa} \left[ -6 \left(\frac{\dot{a}}{a}\right)^2 + \frac{1}{2} \frac{\dot{\xi}^2 + \dot{\eta}^2}{\eta^2} - 2\Lambda + {}^{(3)}R \right]; \quad {}^{(3)}R = \frac{1}{a^2} \left\{ -4\frac{a''}{a} + 2\left(\frac{a'}{a}\right)^2 - \frac{1}{2}\frac{\xi'^2 + \eta'^2}{\eta^2} \right\}, \tag{3.3}$$

where ($'$) denotes the spatial derivative with respect to $z$. One can show that the system (3.3) has the dynamical mode of GW. By taking a week field limit $|\xi| \ll 1$, $|\eta - 1| \ll 1$, the variables $(\xi, \eta - 1)$ represent $\otimes$- $\oplus$-mode of GW propagating in the $z$-direction, respectively. The transversality of GW is guaranteed by the momentum constraint.

Now, we shall replace the variables $(\xi, \eta)$ as $\sqrt{2\kappa}(\phi, \sigma)$. The Lagrangian density is rewritten in the form

$$\mathcal{L}_G = \mathcal{L}_{\hat{G}} + \mathcal{L}_{\hat{M}}; \tag{3.4}$$

$$\mathcal{L}_{\hat{G}} = \frac{a^3}{2\kappa} \left[ -6\left(\frac{\dot{a}}{a}\right)^2 + \hat{R}(a) - 2\Lambda \right]; \quad \hat{R}(a) = \frac{1}{a^2} \left\{ -4\frac{a''}{a} + 2\left(\frac{a'}{a}\right)^2 \right\},$$

$$\mathcal{L}_{\hat{M}} = -\frac{a^3}{2} G^{AB} \partial_\mu \Phi_A \partial^\mu \Phi_B; \quad G^{AB} = \frac{1}{2\kappa} \sigma^{-2} \delta^{AB}.$$

This system is equivalent to a conformally flat cosmology with the spacetime metric

$$d\hat{s}^2 = -dt^2 + \hat{\gamma}_{ij} dx^i dx^j; \quad \hat{\gamma}_{ij} = a^2 \delta_{ij}, \tag{3.5}$$

and a scalar field $\Phi_A = (\phi(t, z), \sigma(t, z))$. So we can use the 2nd order GE to analyze the GW system (3.4).

### B. GE analysis

We are in the position to analyze feature of GW by gradient expansion. We assume the generating functionals for the system (3.3) as

$$\mathcal{S}^{(0)} = -\frac{2}{\kappa} \int d^3x a^3 H(\xi, \eta), \tag{3.6}$$

$$\mathcal{S}^{(2)} = \int d^3x a^3 \left[ J\hat{R} + L\xi_{;m}\xi^{;m} + M\xi_{;m}\eta^{;m} + N\eta_{;m}\eta^{;m} \right], \tag{3.7}$$

where the variables $J, L, M, N$ depend on $(\xi, \eta)$. We should keep in mind that $(;)$ and $\hat{R}(a)$ represent the covariant derivative and 3-curvature associated with the 3-metric $\hat{\gamma}_{ij} \equiv a^2 \delta_{ij}$. Hamiltonian and momentum constraint become

$$\mathcal{H} = \frac{2\kappa}{a^3} \left\{ -\frac{1}{24} \left( \frac{\delta \mathcal{S}}{\delta a} \right)^2 a^2 + \frac{1}{2} \left( \frac{\delta \mathcal{S}}{\delta \xi} \right)^2 \eta^2 + \frac{1}{2} \left( \frac{\delta \mathcal{S}}{\delta \eta} \right)^2 \eta^2 \right\}$$
$$+ \frac{a^3}{2\kappa} \left( 2\Lambda - \hat{R}(a) + \frac{1}{2} \frac{\xi'^2 + \eta'^2}{a^2 \eta^2} \right) = 0. \tag{3.8}$$

$$\mathcal{H}_z = \frac{\delta \mathcal{S}}{\delta a} a' + \frac{\delta \mathcal{S}}{\delta \xi} \xi' + \frac{\delta \mathcal{S}}{\delta \eta} \eta' - \frac{1}{3} \left( \frac{\delta \mathcal{S}}{\delta a} a \right)' = 0. \tag{3.9}$$

The evolution equations are given by

$$\frac{\dot{a}}{a} = -\frac{\kappa}{6a^2} \frac{\delta \mathcal{S}}{\delta a}, \quad \dot{\xi} = \frac{2\kappa}{a^3} \eta^2 \frac{\delta \mathcal{S}}{\delta \xi}, \quad \dot{\eta} = \frac{2\kappa}{a^3} \eta^2 \frac{\delta \mathcal{S}}{\delta \eta}. \tag{3.10}$$

Recall that momentum constraint (3.9) is satisfied by the ansatz of the generating functional. We only have to concentrate on the Hamiltonian constraint. The explicit forms of the 0th order and the 2nd order Hamiltonian constraint are shown in Appendix A.

Our first task is to obtain the 0th order solution. There are two type of the solution: type A and type B. The type A solution we shall analyze specifically here could be constructed if we assume that $H, J, L, M, N$ depend only on $\eta$ (For type B, see Appendix C). Then we observe $M$ identically vanishes. The 0th order Hamiltonian constraint (A3) yields

$$H(\eta) = \frac{H_0}{2}(\eta^{\sqrt{3}/2} + \eta^{-\sqrt{3}/2}) \quad ; \quad H_0 \equiv \sqrt{\frac{\Lambda}{3}} \quad (1 \leq \eta \leq +\infty). \tag{3.11}$$

Substituting (3.6) into (3.10) with a help of (A1), we get the evolution equation for the 0th order:

$$\frac{\dot{a}}{a} = H, \quad \dot{\xi} = 0, \quad \dot{\eta} = -4\eta^2 H_\eta. \tag{3.12}$$

From (3.11)(3.12), we obtain the 0th order solution:

$$a^{(0)} = \tilde{a}(z) \left[\sinh 3H_0(t + t_0(z))\right]^{1/3},$$
$$\xi^{(0)} = \tilde{\xi}(z), \qquad (3.13)$$
$$\eta^{(0)} = \left[\tanh \frac{3H_0}{2}(t + t_0(z))\right]^{-\frac{2}{\sqrt{3}}},$$

where $t_0(z)$, $\tilde{a}(z)$, $\tilde{\xi}(z)$ are integration constants. Behavior of this solution in $(\xi, \eta)$ space is illustrated in Fig.1. The meanings of the integration constants are as follows. $t_0$ represents arbitrariness of the time-slice and $\tilde{a}$ means inhomogeneity of the conformal mode. The remained function $\tilde{\xi}$ expresses the inhomogeneity of GW itself. They become the source terms of the 2nd order evolution equations. As we are interested in inhomogeneity by GW itself, we shall not consider $t_0$ and $\tilde{a}$ hereafter. Note that if we drop the spatial dependence in this solution, it belongs to Bianchi type I homogeneous cosmological model. Its asymptotic behavior becomes

$$H_0 t \ll 1 \text{ (Kasner-like expansion)} \cdots a \sim t^{1/3}, \quad \xi = \tilde{\xi}, \quad \eta \sim t^{-\frac{2}{\sqrt{3}}}.$$
$$H_0 t \gg 1 \text{ (de Sitter-like expansion)} \cdots a \sim e^{H_0 t}, \quad \xi = \tilde{\xi}, \quad \eta \sim 1.$$

Now putting $\tilde{a}(z) = 1$, $t_0(z) = 0$ and noting that $\eta$ does not contain a spatial function, we derive the evolution equations up to the 2nd order from (3.10) and (A2):

$$\frac{\dot{a}}{a} = H - \frac{\kappa}{6a^2} L(\xi')^2,$$
$$\dot{\xi} = -\frac{4\kappa}{a^2} L \eta^2 (\xi''), \qquad (3.14)$$
$$\dot{\eta} = -4H_\eta \eta^2 + \frac{2\kappa}{a^2} L_\eta \eta^2 (\xi')^2.$$

When we solve these equations, we replace $\xi', \xi''$ with 0th order quantity $\tilde{\xi}', \tilde{\xi}''$. We should pay attention that the evolution equations for the 2nd order include the non-linear term $(\xi')^2$ except for the linear term $(\xi'')$. The structure of the equations shows that the non-linear terms affects the evolution of conformal mode $a$ and one of the GW mode, $\eta$. Hence two dynamical mode of GW $(\xi, \eta)$ cannot evolve independently and couples each other. We can understand such the effect from the schematic picture, Fig.2. The left-right arrows on the trajectory denotes polarization of the GW in $(\xi, \eta)$-space. When we set its polarization orthogonal to the trajectory at an initial time, as the universe expands, the arrow becomes to incline due to the non-linear coupling of source term and it is no longer orthogonal to the trajectory.

The next task is to solve the evolution equations (3.14). Assuming that the contribution of the spatial derivative terms is small, we can linearize the evolution equations:

$$a = a^{(0)}(1 + \delta a), \quad \xi = \tilde{\xi}(z) + \delta\xi, \quad \eta = \eta^{(0)} + \delta\eta, \tag{3.15}$$

where $\delta a, \delta\xi, \delta\eta$ are perturbed quantities induced by the spatial gradient terms. Substituting them into (3.14), we get the linearized equations. The solution can be written easily in the integral forms:

$$\begin{aligned}
\delta\xi &= -4\kappa \int_0^t dt \frac{L\eta^2}{a^2}(\tilde{\xi}''), \\
\delta\eta &= 2\kappa\dot{\eta} \int_0^t dt \frac{L_\eta \eta^2}{a^2 \dot{\eta}}(\tilde{\xi}')^2, \\
\delta a &= \int_0^t dt \left[ H_\eta \delta\eta - \frac{\kappa L}{6a^2}(\tilde{\xi}')^2 \right],
\end{aligned} \tag{3.16}$$

where we choose the integration constant so that the perturbation vanishes at $t = 0$. In order to analyze them, we need $L$. The second equation of the 2nd order Hamiltonian constraint (A4) becomes

$$HL - 4\eta^2 H_\eta L_\eta = -\frac{1}{4\kappa}\eta^{-2}. \tag{3.17}$$

Then $L$ is written in the integral form as

$$L = \frac{1}{4\sqrt{3}\kappa H_0} \left( \eta^{\sqrt{3}/2} - \eta^{-\sqrt{3}/2} \right)^{\frac{1}{3}} \int_\infty^\eta \frac{d\eta}{\eta^3} \left( \eta^{\sqrt{3}/2} - \eta^{-\sqrt{3}/2} \right)^{-\frac{4}{3}}, \tag{3.18}$$

where we set integration constants so that $L$ vanishes at initial singularity $t \to 0 (\eta \to \infty)$. We estimate the asymptotic behavior of $L$ at $H_0 t \ll 1$ and $H_0 t \gg 1$:

$$L \simeq \begin{cases} -\frac{1}{8(\sqrt{3}+1)\kappa H_0}\eta^{-2-\frac{\sqrt{3}}{2}} \propto -t^{\frac{4}{\sqrt{3}}+1}, & (H_0 t \ll 1) \\ \\ -\frac{1}{4\kappa H_0} + \frac{L_\infty}{4\kappa H_0}\left(\eta^{\sqrt{3}/2} - 1\right)^{1/3}, & (H_0 t \gg 1) \end{cases} \tag{3.19}$$

where $L_\infty$ is a numerical constant defined by

$$L_\infty = \frac{3}{2g(1)} \int_1^\infty g''(s)(s-1)^{2/3} \simeq 0.28 \; ; \quad g(s) \equiv \frac{1}{s^{\frac{4}{\sqrt{3}}-\frac{1}{3}}(s+1)^{\frac{4}{3}}}.$$

Using the result (3.19), asymptotic behavior of perturbation can be evaluated:

1. $H_0 t \ll 1$ (Kasner regime) :

$$\delta\xi \propto +t^{\frac{4}{3}}(\tilde{\xi}''), \quad \delta\eta \propto +t^{\frac{4}{3}+\frac{2}{\sqrt{3}}}(\tilde{\xi}')^2, \quad \delta a \propto +t^{\frac{4}{3}+\frac{4}{\sqrt{3}}}(\tilde{\xi}')^2. \tag{3.20}$$

The result shows that all modes of perturbation grow in time. The growth rate of $\delta\xi$ is the same as the result of the usual cosmological perturbation (we will see next section). However $\delta\eta$, $\delta a$ grow as a purely non-linear effect because they evolve through the non-linear source term by the 0th order inhomogeneity($(\tilde{\xi}')^2$). (3.20) indicates that non-linear effect such as the mode-coupling will play a significant role in the anisotropic universe.

2. $H_0 t \gg 1$ (de Sitter regime) : The perturbations continue to grow by the time $H_0 t_* \sim 1$. When the Kasner regime ends, they reach certain values $\delta a_*, \delta\xi_*, \delta\eta_*$. Then the universe begins de Sitter expansion ($\eta \to 1$) and the perturbations become

$$\begin{aligned}
\delta\xi &\simeq \delta\xi_* - \frac{1}{2H_0^2}\left[(\eta^{\sqrt{3}/2}-1)^{2/3}\right]_{t_*}^t (\tilde{\xi}''), \\
\delta\eta &\simeq \delta\eta_* + \frac{L_\infty}{12\sqrt{3}H_0^2}(\eta^{\sqrt{3}/2}-1)\left[\frac{1}{\eta^{\sqrt{3}/2}-1}\right]_{t_*}^t (\tilde{\xi}')^2, \\
\delta a &\simeq \delta a_* - \frac{1}{48H_0^2}\left[(\eta^{\sqrt{3}/2}-1)^{2/3}\right]_{t_*}^t (\tilde{\xi}')^2.
\end{aligned} \tag{3.21}$$

In this regime, the perturbations cease to grow and approach constant values. The perturbations caused by non-linear source term cannot grow due to rapid cosmic expansion.

To confirm this analysis, we also solved eq.(3.16) numerically. The results are shown in Fig 3a. The lines in the figures represent the evolution of amplitude of GW in gauge-invariant expressions (For the definition of $\Phi, \Psi$, see eq.(4.5)). The solid line is perturbation which comes from the linear source term, the dashed line from the non-linear source term. We can observe that all amplitude of GW become frozen after they grow by $H_0 t* \sim 1$.

We comment on the validity of the approximation before closing this section. The validity condition can be obtained from evolution equations (3.10).

$$\left|\frac{\delta \mathcal{S}^{(0)}}{\delta a}\right| \gtrsim \left|\frac{\delta \mathcal{S}^{(2)}}{\delta a}\right|, \quad \left|\frac{\delta \mathcal{S}^{(0)}}{\delta \eta}\right| \gtrsim \left|\frac{\delta \mathcal{S}^{(2)}}{\delta \eta}\right|.$$

Roughly estimating, they become

$$H^2 \gtrsim \left(\frac{1}{\text{wavelength of GW}}\right)^2. \tag{3.22}$$

Thus, we recognize GE analysis is valid when the scale of inhomogeneity exceeds the Hubble horizon. Moreover eq.(3.22) is always satisfied if it is satisfied at $H_0 t_* \sim 1$(beginning of inflation).

## IV. COMPARISON WITH LINEAR PERTURBATION THEORY

We saw in Sec.III that two dynamical modes of GW $(\xi, \eta)$ does couple through the non-linear source term and their coupling plays an important role at the stage of anisotropic expansion. In this section, in order to clarify its feature further, we shall analyze the linearized GW and compare with the results of GE analysis. Because we want to extract the physically meaningful modes of GW in our discussion, we examine in terms of the gauge-invariant expression.

We start with the definition of the gauge-invariant quantities and derive the gauge-invariant wave equations. In anisotropic metric

$$ds^2 = -dt^2 + b^2 dz^2 + a^2 \left[\frac{\xi^2 + \eta^2}{\eta} dx^2 + 2\frac{\xi}{\eta} dxdy + \frac{1}{\eta} dy^2\right], \tag{4.1}$$

we will define the perturbed quantities as follows:

$a \to a_0(t)\left(1 + \delta a(t,z)\right), \quad b \to a_0(t)\left(1 + \delta b(t,z)\right), \quad \xi \to \xi_0(t) + \delta\xi(t,z), \quad \eta \to \eta_0(t) + \delta\eta(t,z),$

(4.2)

where subscript ($_0$) denotes the homogeneous background solutions. Note that the metric (4.1) includes the residual variable $b$ which does not appear in (3.1). It is necessary for deriving the physically meaningful equations. Consider the gauge transformation $x^\mu \to \hat{x}^\mu = x^\mu + \varepsilon^\mu$ ($\mu = t, z$) which preserves plane-symmetry. Under the synchronous time slicing, the solution of $\varepsilon^\mu$ can be written containing arbitrarily spatial functions $\chi(z), f(z)$

$$\varepsilon_t = -\chi(z), \quad \varepsilon_z = \chi'(z) a_0^2 \int dt\, a_0^{-2} + f(z). \tag{4.3}$$

This arbitrariness produces the gauge-dependent modes in the metric perturbation. The perturbed quantities are transformed as

$$\begin{aligned}
\delta \tilde{a} &= -H\chi, \\
\delta \tilde{\xi} &= -\dot{\xi}_0 \chi, \\
\delta \tilde{\eta} &= -\dot{\eta}_0 \chi, \\
\delta \tilde{b} &= -H\chi - \chi'' \int \frac{dt}{a_0^2} - f'/a_0^2,
\end{aligned} \tag{4.4}$$

where $H = \dot{a}_0/a_0$. It is obvious that every quantities containing $\delta b$ become gauge-dependent. Using this result, we define the gauge-invariant quantities $\Phi, \Psi$:

$$\Phi = (\delta\eta + \frac{\xi_0}{\eta_0}\delta\xi) - \frac{1}{H}\left(\frac{\xi_0 \dot{\xi}_0 + \eta_0 \dot{\eta}_0}{\eta_0}\right)\delta a, \quad \Psi = (\delta\xi - \frac{\xi_0}{\eta_0}\delta\eta) - \frac{1}{H}\left(\dot{\xi}_0 - \frac{\dot{\eta}_0}{\eta_0}\xi_0\right)\delta a. \tag{4.5}$$

We derive the gauge-invariant perturbed equations. In the case of type A background solution

$$\xi_0 = 0, \quad \eta_0 = \left[\tanh\frac{3H_0 t}{2}\right]^{-\frac{2}{\sqrt{3}}}, \tag{4.6}$$

then the gauge-invariant quantities can be reduced as follows:

$$\Phi = \delta\eta - \frac{\dot{\eta}_0}{H}\delta a, \quad \Psi = \delta\xi. \tag{4.7}$$

In order to derive the gauge-invariant equations, we use perturbed Einstein equations represented in Appendix B. The evolution equation for $\Phi$ is obtained from (B2) (B4)(B6) and $\Psi$ obtained from (B3):

$$\ddot{\Phi} + (3H - 2\frac{\dot{\eta}_0}{\eta_0})\dot{\Phi} + \left(\frac{\dot{\eta}_0}{\eta_0}\right)^2 \left\{1 + \frac{3}{2}\left(\frac{H_0}{H}\right)^2\right\}\Phi - \Phi''/a_0^2 = 0, \qquad (4.8)$$

$$\ddot{\Psi} + (3H - 2\frac{\dot{\eta}_0}{\eta_0})\dot{\Psi} - \Psi''/a_0^2 = 0. \qquad (4.9)$$

Eq.(4.8) and (4.9) are decoupled so that one can solve them independently. In other words, a suitable choice of "polarization" does not make the mode-coupling in linear GW. When we substitute (3.16) into (4.7), one can easily see that $\Phi$, $\Psi$ in GE analysis no longer evolve independently because of the non-linear source term. Therefore we have confirmed that mode-coupling of GW that appeared in the analysis by GE can be recognized as a purely non-linear effect.

Now, let us compare the growing mode obtained from both analysis using the gauge-invariant quantities $\Phi, \Psi$. As for GE analysis, we may directly substitute the results (3.16) into (4.7). When we evaluate the linear perturbation(LP) from (4.8)(4.9), we take the long-wavelength limit($k/H_0 \lesssim 1$) in terms of the Fourier expansion. The results in Kasner regime are represented in Table I. One can see that the LP contains the same growing mode that is generated from the linear source term in GE analysis. When $H_0 t \gg 1$, both modes of the LP become the same equation as

$$\ddot{\Phi} + 3H_0 \dot{\Phi} = 0. \qquad (4.10)$$

We can observe from this equation that the quantities $\Phi$, $\Psi$ obey the same equation of motion as GW in FRW universe, and all growing modes become frozen in the long-wavelength limit. This behavior is the same as (3.21). Accordingly the perturbation from the linear source term in GE is consistent with the result of LP in the long-wavelength limit. We may also recognize this thing numerically (Fig.3b). The solid, dashed and other lines represent the amplitude $\Psi$ of LP for the various wave number $k/H_0$. After solving numerically with the initial condition $\dot{\Psi} = 0$, we shift the curves so that they start at the origin. In long-wavelength limit ($k/H_0 \lesssim 1$), the LP coincides with the markers, which is the perrturbation from the linear source term in GE.

| type A | $\Phi$ | $\Psi$ |
|---|---|---|
| LP | $t^{-\frac{2}{\sqrt{3}}} J_0(\frac{3k}{2H_0}(H_0 t)^{2/3}) \sim t^{-\frac{2}{\sqrt{3}}}(c_0 + c_2 t^{\frac{4}{3}} + \cdots)$ | $t^{-\frac{2}{\sqrt{3}}} J_{\sqrt{3}}(\frac{3k}{2H_0}(H_0 t)^{2/3}) \sim d_0 + d_2 t^{\frac{4}{3}} + \cdots$ |
| GE | $t^{\frac{4}{\sqrt{3}}+\frac{2}{3}}(\tilde{\xi}')^2$ (non-linear) | $t^{\frac{4}{3}}(\tilde{\xi}'')$ (linear) |

TABLE I. comparison of the growing mode of GW (type A)

As for type B solution, similar discussion leads us to the same result as type A: Substituting the homogeneous solution of (C3) into (4.5), (B2)∼(B4) yield

$$\ddot{\Phi} + 3H\dot{\Phi} - \dot{\phi}^2 \Phi - \Phi''/a_0^2 = 0, \quad (4.11)$$

$$\ddot{\Psi} + 3H\dot{\Psi} - \frac{1}{2}\dot{\phi}\left(\frac{\dot{\phi}}{H}\right)^{\cdot} \Psi - \Psi''/a_0^2 = 0. \quad (4.12)$$

There is also no evidence of the mode-coupling in linearized GW for type B.

## V. SUMMARY AND DISCUSSION

In this paper, we have considered the inhomogeneous cosmological model including GW in terms of GE analysis. Imposing plane-symmetry, dynamical modes of GW can be replaced into that of scalar fields, hence it has become possible to analyze the feature of GW by using the present expansion scheme. The result show us that a non-linearity is incorporated as an effect of mode-coupling and it plays an important role at the stage of anisotropic expansion. We have also confirmed that the GW generated from the linear source term in GE is same growth as the result of linear perturbation theory.

Consider how initial inhomogeneities affects the present universe. During the inflationary regime, the scale of the initial inhomogeneity becomes larger than the Hubble horizon. After the end of inflation, since the horizon becomes larger, the scale of the inhomogeneity reenters the horizon sooner or later. We can observe such the inhomogeneity as fluctuation of CMB due to the fluctuation of the gravitational potential (Sachs-Wolfe effect [14]). In the presence of polarized of GW initially, our model indicates that the photon of CMB might be polarized through the non-linear evolution. Taking into account of this non-linearity, we can in principle get the information of the early universe from the measurement of CMB. Because our model is just preliminary one, we need to study the more realistic model further to compare with the observation precisely.

In fact, we should remark that our treating model belongs to the particular case of Bianchi type I so that GW can propagate only to the z-direction (see Sec.III A). Thus, in order to obtain more general consequence and develop the theoretical prediction to the observations, it is necessary to extend our analysis to more general case. But we have not succeeded in applying the present expansion scheme to such a general case because we cannot treat GW as matter field in such a case. In general anisotropic case, GW has a gauge-dependent mode-coupling even in linear level and we cannot have any suitable gauge to eliminate this coupling [15]. We must consider a new formulation of the expansion scheme. In order to clarify inhomogeneous cosmology, we will proceed to study about this formulation as a next step [16].

## ACKNOWLEDGMENTS

We would like to thank Prof.Tomimatsu, Dr.Nakamura for critical comments and valuable discussions. One of us (A.T.) would also like to thank Dr.Yamaguti for useful communications and encouragement.

## APPENDIX A: HAMILTON-JACOBI EQUATIONS AND VARIATIONAL FORMULAE

In this appendix, we shall derive the 0th and 2nd order Hamiltonian constraint .

First, from the ansatz of the generating functional form (3.6)(3.7), we obtain the following variational formulae:

$$\frac{\delta \mathcal{S}^{(0)}}{\delta a} = -\frac{6}{\kappa} a^2 H, \quad \frac{\delta \mathcal{S}^{(0)}}{\delta \xi} = -\frac{2}{\kappa} a^3 H_\xi, \quad \frac{\delta \mathcal{S}^{(0)}}{\delta \eta} = -\frac{2}{\kappa} a^3 H_\eta. \tag{A1}$$

$$\frac{\delta \mathcal{S}^{(2)}}{\delta a} = a^2 \left[ J\hat{R} + L(\xi_{;m}\xi^{;m}) + M(\xi_{;m}\eta^{;m}) + N(\eta_{;m}\eta^{;m}) - 4J_{;m}{}^{;m} \right],$$

$$\frac{\delta \mathcal{S}^{(2)}}{\delta \xi} = a^3 \left[ J_\xi \hat{R} - L_\xi(\xi_{;m}\xi^{;m}) - 2L_\eta(\xi_{;m}\eta^{;m}) + (N_\xi - M_\eta)\eta_{;m}\eta^{;m} - 2L\xi_{;m}{}^{;m} - M\eta_{;m}{}^{;m} \right], \quad \text{(A2)}$$

$$\frac{\delta \mathcal{S}^{(2)}}{\delta \eta} = a^3 \left[ J_\eta \hat{R} + (L_\eta - M_\xi)\xi_{;m}\xi^{;m} - 2N_\xi(\xi_{;m}\eta^{;m}) - N_\eta(\eta_{;m}\eta^{;m}) - M\xi_{;m}{}^{;m} - 2N\eta_{;m}{}^{;m} \right],$$

where the covariant derivative (;) is also induced by the 3-metric $\hat{\gamma}_{ij} = a^2 \delta_{ij}$.

Using these formulae, Hamiltonian constraint (3.8) can be expanded by spatial gradient. The resulting equations are

$$\mathcal{H}^{(0)} = 0 \quad \cdots \quad 3H^2 - \Lambda = 4\eta^2(H_\xi^2 + H_\eta^2). \tag{A3}$$

$$\mathcal{H}^{(2)} = 0 \quad \cdots \quad \begin{cases} HJ - 4\eta^2(H_\xi J_\xi + H_\eta J_\eta) = \frac{1}{2\kappa} \\[6pt] H(L - 4J_{\xi\xi}) - 4\eta^2 H_\eta(L_\eta - M_\xi) + 4\eta^2 H_\xi L_\xi = -\frac{1}{4\kappa}\eta^{-2} \\[6pt] H(N - 4J_{\eta\eta}) - 4\eta^2 H_\xi(N_\xi - M_\eta) + 4\eta^2 H_\eta N_\eta = -\frac{1}{4\kappa}\eta^{-2} \\[6pt] H(M - 8J_{\xi\eta}) + 8\eta^2(H_\xi L_\eta + H_\eta N_\xi) = 0 \\[6pt] HJ_\xi = \eta^2(2H_\xi L + H_\eta M) \\[6pt] HJ_\eta = \eta^2(2H_\eta N + H_\xi M) \end{cases} \tag{A4}$$

At first glance, One can observe that number of equations exceeds that of unknown variables and they seem to be inconsistent. However, two equations of (A4) are trivial due to internal relation. there is no contradiction (see Ref. [7]). When we solve them concretely for type A or B, we can further proceed to reduce them to much simple forms. As for type A, we assume $H, J, K, L, M, N$ depend only on $\eta$. Then $M$ vanishes identically and (A4) can be reduced to 4 equations.

## APPENDIX B: PERTURBED EINSTEIN EQUATIONS

Here, we write down the perturbed Einstein equations under the metric (4.1),

$$ds^2 = -dt^2 + b^2 dz^2 + a^2 \left[ \frac{\xi^2 + \eta^2}{\eta} dx^2 + 2\frac{\xi}{\eta} dx dy + \frac{1}{\eta} dy^2 \right].$$

Using the perturbed quantities defined in eq.(4.2), the perturbed equations of motion are obtained:

$$\delta\ddot{a} + \delta\ddot{b} + 3H(\delta\dot{a} + \delta\dot{b}) + \frac{1}{2\eta_0^2}\left\{\dot{\xi}_0\delta\dot{\xi} + \dot{\eta}_0\delta\dot{\eta} - \left(\frac{\dot{\xi}_0^2 + \dot{\eta}_0^2}{\eta_0}\right)\delta\eta\right\} = \frac{\delta a''}{a_0^2}, \tag{B1}$$

$$\delta\ddot{a} + H(5\delta\dot{a} + \delta\dot{b}) = \frac{\delta a''}{a_0^2}, \tag{B2}$$

$$\delta\ddot{\xi} + (3H - 2\frac{\dot{\eta}_0}{\eta_0})\delta\dot{\xi} + \frac{\dot{\xi}_0}{\eta_0}\left(2\frac{\dot{\eta}_0}{\eta_0}\delta\eta - 2\delta\dot{\eta}\right) + \dot{\xi}_0(2\delta\dot{a} + \delta\dot{b}) = \frac{\delta\xi''}{a_0^2}, \tag{B3}$$

$$\delta\ddot{\eta} + (3H - 2\frac{\dot{\eta}_0}{\eta_0})\delta\dot{\eta} + (\frac{\dot{\eta}_0^2 - \dot{\xi}_0^2}{\eta_0^2})\delta\eta + 2\frac{\dot{\xi}_0}{\eta_0}\delta\dot{\xi} + \dot{\eta}_0(2\delta\dot{a} + \delta\dot{b}) = \frac{\delta\eta''}{a_0^2}, \tag{B4}$$

where $H = \dot{a}_0/a_0$, $a_0 = [\sinh H_0 t]^{1/3}$. In addition, we have two constraint equations of perturbation corresponding to the Hamiltonian and momentum constraint:

$$H(2\delta\dot{a} + \delta\dot{b}) - \frac{1}{4\eta_0^2}\left\{\dot{\xi}_0\delta\dot{\xi} + \dot{\eta}_0\delta\dot{\eta} - \left(\frac{\dot{\xi}_0^2 + \dot{\eta}_0^2}{\eta_0}\right)\delta\eta\right\} = \delta a''/a_0^2, \tag{B5}$$

$$-\delta\dot{a}' = \frac{\dot{\xi}_0\delta\xi' + \dot{\eta}_0\delta\eta'}{4\eta_0^2}. \tag{B6}$$

## APPENDIX C: THE TYPE B SOLUTION

We shall investigate another solution of the system (3.3), so called type B. The type B solution up to the 2nd order can be obtained by the following assumption in the functional form :

$$H = H(x), \quad J = J(x), \quad L = \eta^{-2}\tilde{L}(x), \quad M = \eta^{-2}\tilde{M}(x), \quad N = \eta^{-2}\tilde{N}(x); \quad x \equiv \xi/\eta. \tag{C1}$$

In similar manner for type A, we start to construct the 0th order solution. Using (C1) and (A3) lead us to the solution $H(x)$ as

$$H(x) = \frac{H_0}{2}(y^{\sqrt{3}/2} + y^{-\sqrt{3}/2}); \quad y = x + \sqrt{x^2 + 1}. \tag{C2}$$

Thus we can get the 0th order time evolution:

$$\xi^{(0)} = C(z)\tanh\phi(t,z), \tag{C3}$$

$$\eta^{(0)} = C(z)\mathrm{sech}\phi(t,z),$$

where

$$\phi(t,z) + \phi_0 = -\frac{2}{\sqrt{3}}\ln\left|\tanh\frac{3H_0}{2}(t+t_0(z))\right|. \tag{C4}$$

The trajectory of this solution is depicted in Fig.1. Note that $a^{(0)}(t,z)$ becomes the same solution as type A ( see eq.(3.13)). This means that the asymptotic behavior of type B can also be classified in two regime: Kasner regime($H_0 t \ll 1$) and de Sitter regime($H_0 t \gg 1$).

We derive the evolution equations up to the 2nd order. The 0th order solution (C3) has the spatial function $C(z)$ which represents the inhomogeneity of GW. When we put $\tilde{a}(z) = 1$, $t_0(z) = 0$, the evolution equations are reduced by using (A4) (after some manipulations):

$$\frac{\dot{a}}{a} = H - \frac{\kappa P(x)}{6a^2}\left(\frac{C'}{C}\right)^2, \tag{C5}$$

$$\begin{pmatrix}\dot{\xi}\\\dot{\eta}\end{pmatrix} = -4\eta^2\begin{pmatrix}H_\xi\\H_\eta\end{pmatrix} - \frac{2\kappa}{a^2}\begin{pmatrix}\boldsymbol{L} & -\boldsymbol{N}\\\boldsymbol{N} & \boldsymbol{L}\end{pmatrix}\begin{pmatrix}\xi\\\eta\end{pmatrix}, \tag{C6}$$

where,

$$P(x) \equiv x^2\tilde{L} + \tilde{N} + x\tilde{M}, \quad \boldsymbol{L} \equiv \frac{2P}{1+x^2}\left(\frac{C'}{C}\right)', \quad \boldsymbol{N} \equiv P_x\left(\frac{C'}{C}\right)^2.$$

$P(x)$ obeys the following equation:

$$4(1+x^2)H_x P_x - HP = \frac{1}{4\kappa}(x^2+1). \tag{C7}$$

Eq.(C6) shows us that the linear source term $\boldsymbol{L}$ appears in the diagonal part of the right hand side matrix, on the other hand, the non-linear term $\boldsymbol{N}$ in off-diagonal part. Because of their time dependencies, it is obvious that two dynamical mode of GW no longer decouple all over the time. From the result in Sec.IV, a purely non-linear effect is also incorporated as the mode-coupling of GW in type B.

We evaluate (C5)(C6). Dividing them into the perturbed and unperturbed part(see eq.(3.15)) and rewrite them in terms of the gauge-invariant expressions (4.5). The results are

$$\Phi_{ge} = \delta\eta + x\delta\xi = -\frac{4\kappa}{\eta}\int_{t_i}^{t}dt\frac{P}{a^2(1+x^2)}\left(\frac{C'}{C}\right)',$$

$$\Psi_{ge} = \delta\xi - x\delta\eta - \frac{\dot\phi}{H}\delta a = \frac{\kappa\dot\phi}{H}\int_{t_i}^{t}dt\left\{\frac{2HP_x(1+x^2)}{a^2\dot x} + \frac{P}{6a^2}\right\}\left(\frac{C'}{C}\right)^2. \quad (C8)$$

We set all perturbation vanishes at $t = t_i(H_0 t_i \ll 1)$. We can solve them numerically. Their behavior is the same as type A qualitatively: they grow in Kasner regime and become frozen in de Sitter phase.

Finally, we comment on the differences between type A and B. For 0th order trajectory in $(\xi, \eta)$-space, type B can be represented by a portion of a circle, while type A draws a straight line pallarel to $\eta$-axis (Fig.1). As a portion of circle in large raudius limit approximates a line, behavior of the type B solution for the 2nd order could also be explained qualitatively from type A.

# Figure Captions

1. Trajectories of two types of the solutions in $(\xi, \eta)$-space: The straight line is the trajectory of type A solution. The curved line which is a portion of circle corresponds to type B solution.

2. Schematic picture of the mode-coupling of GW in $(\xi, \eta)$-space: (a)type A; (b)type B. Left-right arrows on each trajectories denotes the polarization of GW in $(\xi, \eta)$-space. Due to the non-linear coupling, the initial arrows orthogonal to the trajectory become to incline and no longer orthogonal at the end of Kasner regime (see eq.(3.14)(C6)).

3. Evolution of gravitational wave for type A solution : Horizontal axis denotes time normalized by $H_0^{-1}$ and vertical axis means the amplitude of GW $\Phi, \Psi$ defined in (4.5).

   (a) $\cdots$ Results of the GE analysis: The solid line represents $\Psi$ which corresponds to perturbation from the linear source term and the dashed line is $\Phi$ from the non-linear source term. Solving $L$ from (3.18) and substituting (3.16) into (4.5), both lines are obtained numerically.

   (b) $\cdots$ Comparison between GE and LP: The markers show evolution from the linear source term in GE which is the same solid curve in (a). On the other hand, the solid, dashed and other lines are the amplitude of linearized GW $\Psi$ for the various wave number $k/H_0$. After solving (4.9) numerically under the initial condition $\dot{\Psi} = 0$, we shift the curves so that they start at the origin.

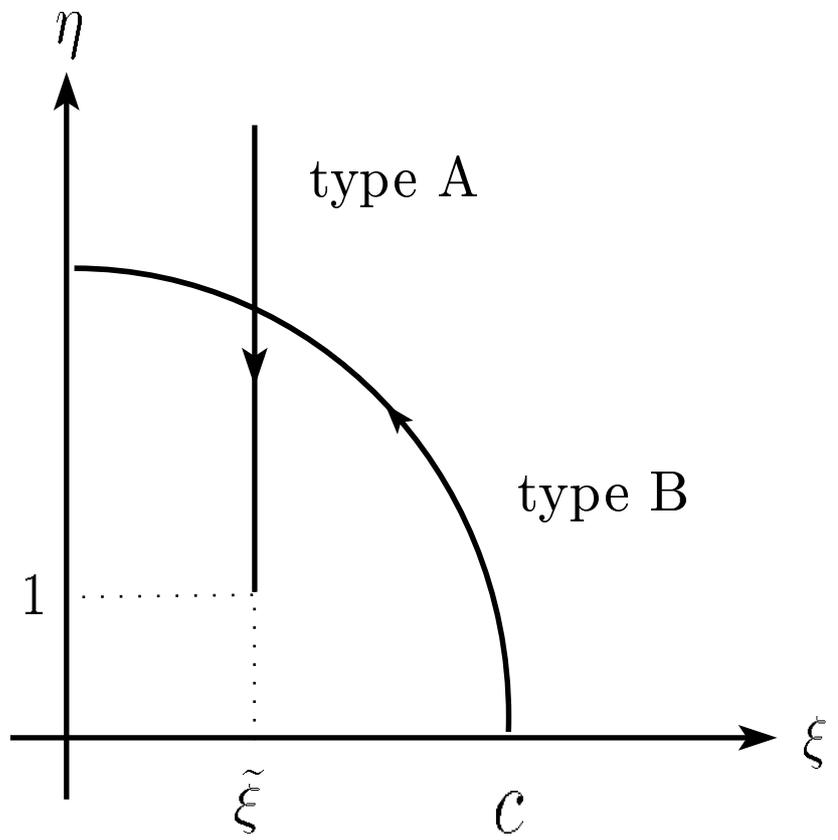

FIG. 1.

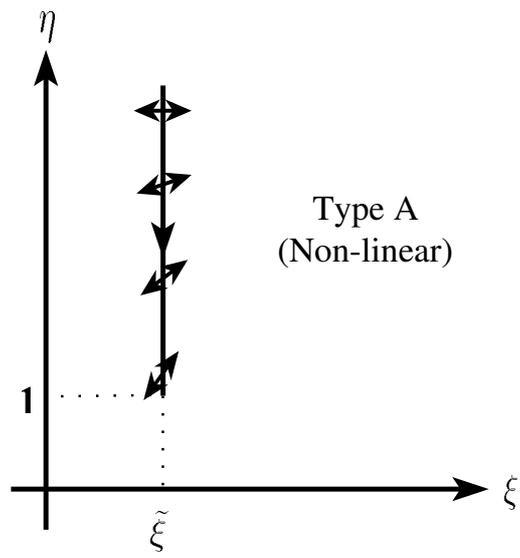

(a)

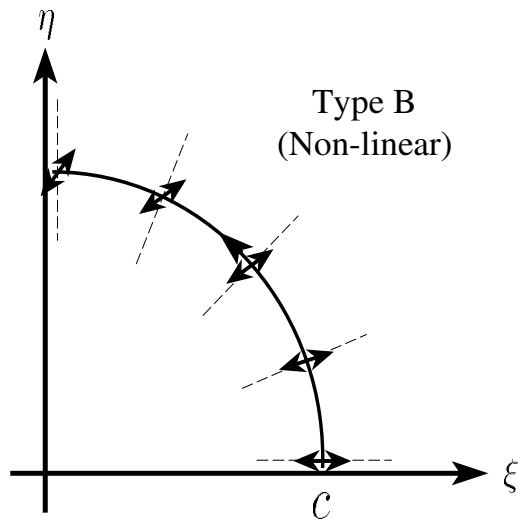

(b)

FIG. 2.

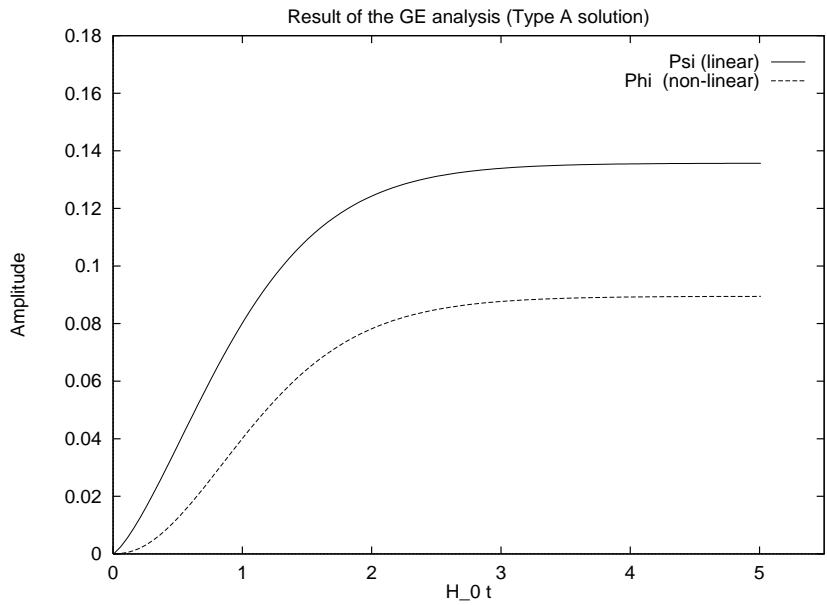

(a)

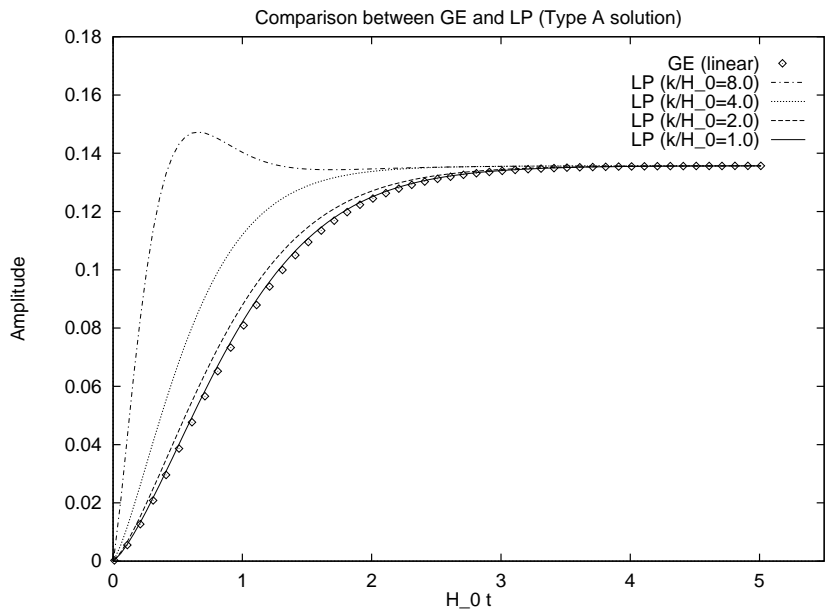

(b)

FIG. 3.